\documentclass[aps,preprintnumbers,amsmath,amssymb,11pt]{revtex4}

\usepackage{epsf}

\usepackage{graphicx}

\usepackage{bm}

\begin{document}


\def\Qs{Q_{\rm s}}

\def\half{{\textstyle{\frac12}}}
\def\p{{\bm p}}
\def\q{{\bm q}}
\def\x{{\bm x}}\def\v{{\bm v}}
\def\E{{\bm E}}
\def\B{{\bm B}}
\def\A{{\bm A}}
\def\a{{\bm a}}
\def\b{{\bm b}}
\def\c{{\bm c}}
\def\j{{\bm j}}
\def\n{{\bm n}}
\def\grad{{\bm\nabla}}
\def\da{d_{\rm A}}
\def\tr{\operatorname{tr}}
\def\Im{\operatorname{Im}}
\def\Re{\operatorname{Re}}
\def\md{m_{\rm D}}
\def\mpl{m_{\rm pl}}
\def\pol{\varepsilon}
\def\bpol{{\bm\pol}}
\def\CP{CP^{N-1}}
\def\be{\begin{equation}}
\def\ee{\end{equation}}
\def\bea{\begin{eqnarray*}}
\def\eea{\end{eqnarray*}}


\title
    {
Bosonization and the Berry connection in two-dimensional QED
    }

\author {H. B. Thacker\footnotemark \footnotetext{email: hbt8r@virginia.edu} and 
Gabriel Wong
}
\affiliation
    {%
 Department of Physics,
    University of Virginia,
    P.O. Box 400714
    Charlottesville, VA 22901-4714\\
}

\date{\today}

\begin {abstract}%
{ 
The dynamical effects of topological charge in two-dimensional QED can be expressed in terms of
a topological order parameter via a Berry phase construction. The Berry phase describes the
electric charge polarization of the vacuum in a manner similar to the theory of polarization in
topological insulators. The topological order parameter labels 
discrete vacua which differ by units of electric flux. Here the associated Berry connection
is explicitly constructed from the Dirac Hamiltonian eigenstates by introducing a small attractive Thirring
coupling, so that there is still a stable boson in the limit of zero EM coupling. The Berry connection arises 
from the analytic structure of the Bethe ansatz states in complex rapidity near the free fermion point. 
}%
\end {abstract}

\maketitle
\thispagestyle {empty}

\newpage
\section {Introduction}

The most well-verified phenomenological effect of topological gauge fluctuations in the QCD vacuum
is the resolution of the U(1) problem (i.e. the absence of a low mass flavor singlet Goldstone boson)
via a topologically induced $\eta'$ mass term. Phenomenology and lattice studies of the $\eta'$ strongly support the Witten-Veneziano
large $N_c$ arguments \cite{Witten_largeN,DiVecchia,Schechter} which predict the $\eta'$ mass in terms of the topological
susceptibility of the pure-glue vacuum. In this view, the $\eta'$ is an approximate Goldstone boson in the 
large $N_c$ limit (with $m_{\eta'}^2\propto 1/N_c$). In this limit, the topologically induced $\q\bar{q}$ annihilation
that provides the flavor singlet mass term is suppressed, and the $\eta'$ propagates like a pion. Both large $N_c$ chiral
Lagrangian arguments \cite{Witten_largeN,DiVecchia,Schechter} and Monte Carlo results \cite{Horvath03,Ilgenfritz07} indicate that the topological 
structure of the QCD vacuum is dominated by 2+1 dimensional membranes of topological charge, which can be understood
as domain walls separating discrete vacua in which the effective local value of the topological $\theta$ parameter differs
by integer multiples of $2\pi$. As a gauge excitation, a $\theta$-domain wall can be inserted in the action as an integral
of the 3-index Chern-Simons tensor over the 2+1 dimensional world volume of the membrane. Lattice results indicate that the vacuum is permeated by a
layered, alternating sign arrangement of these topological charge membranes. The polarizability of this stack of membranes 
leads to the finite topological susceptibility of the QCD vacuum. 

Recently, it has been suggested \cite{Thacker14} that a natural mathematical
framework for discussing the topological structure of the QCD vacuum and its relation to the chiral condensate is provided by the theory of 
electric polarization and quantized charge transport in topological insulators \cite{Vanderbilt}. The central idea in the theory of topological insulators
is the construction of a Berry connection which is associated with the phase of Bloch wave fermion eigenstates under adiabatic
transport around a compact momentum-space Brillouin zone. The Berry connection is thus a gauge field defined over momentum
space rather than coordinate space. The Berry phase given by the closed Wilson loop integral around the BZ is gauge invariant under 
small, topologically trivial gauge variations of the Berry connection, but it changes by integer multiples of $2\pi$ under topologically nontrivial transformations,
corresponding to the transfer of integer units of charge between the two spatial boundaries of the system. The Berry phase provides a
definition of electric polarization which properly incorporates quantized charge transport in a topological insulator or quantum Hall system.

A very instructive analogy to large $N_c$ vacuum structure in QCD is provided by $U(1)$ electrodynamics in two spacetime
dimensions. Here the Chern-Simons membrane is constructed from an ordinary Wilson line integral of the gauge field (interpreted as the surface integral
of the dual Chern-Simons current $\varepsilon_{\mu\nu}A^{\nu}$). Thus the domain walls are world lines of charged particles and the discrete $\theta$-vacua
are flux vacua with different numbers of units of electric flux. This provides some insight into the physical significance 
of the Berry connection. In one spatial dimension, there is a direct connection between the ordering of particles along the
spatial axis and the analytic structure of amplitudes and wave functions in complex momentum space. For example, a 2-body
wave function $\Psi(x_1,x_2)$ that vanishes for one ordering of the particles (e.g. $x_1<x_2$) gives a momentum space wave function
that is analytic in the upper or lower half-plane of the relative momentum variable $k=k_1-k_2$. In the discussion of the Berry phase
in Ref. \cite{Thacker14}, gauge topology was related to spectral flow for the free massless Dirac Hamiltonian in a background gauge field.
However, for QED2 the analytic structure of the Berry connection becomes more clear if we introduce a small Dirac mass term as well as a small Thirring
4-fermion coupling. Thus we are led to consider the spectral flow of the massive Thirring Hamiltonian in a constant background EM field strength.
Without the EM field, this is is still a solvable model by a Bethe ansatz \cite{Bergknoff79,Thacker81}.
The construction of its particle spectrum is given in terms of Hamiltonian eigenstates described by a single quasiparticle energy function in complex momentum space (or more conveniently,
complex rapidity space). In particular, the elementary boson of the equivalent sine Gordon model \cite{Coleman75} is a fermion-antifermion
bound state associated with a pole in the analytically continued fermion-antifermion phase shift. For any small but finite Thirring coupling $g>0$ (giving an
attractive fermion-antifermion interaction), the SG boson is a well-defined
particle, but this state dissociates at the free fermion point due to the coalescence of a pole and a zero in the 2-body amplitude.
For this reason, the analytic structure is made clearer by approaching the free fermion point from a small positive coupling.
Below we briefly review the relevant spectral results for the massive Thirring model near the free fermion point. Our 
construction of the Berry connection is built on the analytic structure of the Bethe ansatz eigenfunctions in complex rapidity
for small 4-fermion coupling $g>0$.

As discussed in \cite{Thacker14}, the physical significance of the Berry phase in QED2 is that it represents the electric polarization,
with mod $2\pi$ jumps of $\theta$ describing the topological transport of charge between boundaries. The equivalence to sine Gordon theory
provides some insight here. The fermion is a kink in the SG field, which changes by $2\pi$ across the kink. (For this discussion, we absorb a factor of
the SG coupling constant $\beta$ in the definition of $\phi$ so that the action is periodic in $\phi\rightarrow \phi+2\pi$.) From a semiclassical perspective, 
a kink-antikink pair that ionizes and propagates to opposite ends of the box leaves an SG field which is not zero but $\pm 2\pi$. If there is no
EM field, $\phi\rightarrow\phi+2\pi$ is an exact symmetry, so all the mod $2\pi$ vacua are degenerate. However, with an EM field the 
symmetry is broken by the Schwinger anomaly. This is most easily seen by bosonizing the theory in Coulomb gauge, where the EM field reduces
to an instantaneous linear Coulomb potential $V(x-y)=|x-y|$. Under bosonization, this becomes a mass term for the boson field,
\begin{equation}
H_{EM}= e^2 \int dx dy j_0(x) V(x-y) j_0(y)\propto  e^2\int dx \phi^2(x)
\end{equation}
where the current is bosonized according to $j_{\mu}\propto \varepsilon_{\mu\nu}\partial^{\nu}\phi$. The vacua in which $\phi$ differs by mod $2\pi$ 
are thus distinct ``flux vacua'' containing different numbers of units of electric flux. This is quite analogous to the situation expected in
large-$N_c$ QCD, where there are discrete vacua for $\theta=2\pi\times integer$, whose degeneracy is broken by a contribution to the vacuum 
energy $\propto \theta^2$ coming from the finite topological susceptibility of the vacuum. Thus we expect the present discussion of the Berry connection in QED2 to have
direct relevance to 4D QCD, where it describes the polarization of Chern-Simons membranes.

Our strategy for constructing the Berry connection in two-dimensional QED follows the discussion in Ref. \cite{Thacker14}. We consider the spectral flow
of the Dirac Hamiltonian in a background U(1) gauge field defined on a Euclidean spacetime 2-torus, with spatial period $2\pi$ and Euclidean time period $T$. In the 2D case,
the topological charge is proportional to the field strength $F=\frac{1}{2}\varepsilon_{\mu\nu}F^{\mu\nu}$. For a constant
background field strength $F$, flux quantization requires $FT=$ integer. With the gauge choice $A_0=0, A_1=Ft$, the one-body Dirac Hamiltonian operator may be regarded as a 
function of the parameter $k\equiv Ft$,
\begin{equation}
\label{eq:spectralflow}
H_0(k)=\gamma^5\left(-i\frac{\partial}{\partial x} + k\right) + \gamma^0 m
\end{equation}
with eigenfunctions $u(x,k)$. If there are other Dirac particles in the state, we consider the many body wave function $\Psi(x_1,\ldots,x_N)$ as a function of $x\equiv x_1$, 
for fixed $x_2,\ldots,x_N$. The Thirring interaction will introduce delta-function interaction terms in (\ref{eq:spectralflow}).
\begin{equation}
H(k) = H_0(k) + 2g\sum_{j=2}^{N}\delta(x-x_j)
\end{equation}
These are treated by writing eigenstates of the free Hamiltonian (\ref{eq:spectralflow}) in each sector and matching them across the delta functions with
two-body phase shifts.
As discussed in \cite{Thacker14}, for $T>>1$ the time-dependent Schrodinger equation for $H$ may be interpreted as an equation for the adiabatic evolution of the Hamiltonian eigenstates
as a function of $k$. The Euclidean time period $0<t<T$ corresponds to $0<k<1$. If we take $H(k)$ to act on a periodic spatial interval $0<x<2\pi$, then 
for any integer value of $k$, the Hamiltonian is gauge equivalent to $H_0(0)$,
\begin{equation}
\label{eq:Hamiltonian}
H(k) = e^{-ikx}H(0)e^{ikx}
\end{equation}
In the $A_0=0$ gauge, the eigenfunctions of $H_0(k)$ on the torus are periodic in the spatial direction,
\begin{equation}
u(x+2\pi,k)=u(x,k),
\end{equation}
but they are only quasiperiodic in the time direction,
\begin{equation}
u(x,k+1) = e^{ix}u(x,k)
\end{equation}
The gauge transformation required to match wave functions at $t=0$ and $t=T$ represents quantized spectral flow of the Hamiltonian
induced by the background $F$ field. For example, in a constant F field $F=1/T$, the eigenvalues of the massless left-handed Dirac Hamiltonian, $E_n=n+k$, evolve continuously
from $E_n=n$ to $E_n=n+1$ as we go from 0 to $T$. The axial vector current anomaly displays the connection between the spectral flow description of gauge
field topology and the polarization of electric charge. In the chiral limit $m\rightarrow0$, in the $A_0=0$ gauge, the nonconservation of axial current is attributed to the time-dependence of 
the $A_1=Ft$ background field,
\begin{equation}
\partial^{\mu} j^5_{\mu} = \partial^0 A_1 = F
\end{equation}
Thus, the quantization of the gauge flux $F$ on a Euclidean torus translates into the quantization of the change of total axial charge on the periodic spatial cell under
evolution over a Euclidean time period $T$. This describes the possibility of producing one or more fermion-antifermion pairs which propagate to opposite spatial boundaries
and change the chiral phase of the vacuum by $\pm 2\pi$. 
Later we will consider Bloch wave states on a spatial lattice of length $L\rightarrow \infty$ 
made up of adjacent quasiperiodic cells. In this case a topological
transition describes the propagation of the fermion and antifermion to the boundaries of neigboring cells of the lattice.
The need for a topologically nontrivial gauge transformation to match up the states at $t=0$ and $t=T$ reflects the fact that
the change of phase of a Hamiltonian eigenstate around the Euclidean time period records the transfer of 
charge to the boundaries. The Berry connection in QED2 encodes phase information from the bulk eigenstates that registers the net transfer
of chiral charge to the boundaries. Here we study this issue in the framework of the eigenstates constructed from the massive Thirring model Bethe ansatz near the free fermion limit.
We will show that a particular excitation that appears in the Bethe ansatz spectrum known as a ``2-string'' represents a transition between vacua which differ by a unit of chiral charge.
When the integral around the Brillouin zone that determines the Berry phase is evaluated by contour integration in the complex rapidity plane, each 2-string contributes a pole
residue of $\pm 2\pi$ (the sign depending on the chirality of the 2-string). In the fermionic formulation of the Thirring model, a 2-string is a fermion-antifermion bound state 
which, in the absence of an EM interaction, has zero energy. It is a scalar excitation with the symmetries of a $\bar{\psi}\psi$ operator. 
Thus the 2-string can be seen as the building block of the chiral condensate. (Note that, although the Coleman-Mermin-Wagner theorem prevents sponaneous chiral symmetry breaking in
2D, the covariant gauge description of QED2 includes a massless Goldstone field which cancels against the massless photon pole and thereby decouples from gauge invariant 
amplitudes.)

Applying the gauge transformation $g(x)=e^{iFxt} =e^{ikx}$ we go to Coulomb gauge, $A_0=-Fx,\;A_1=0$. In this gauge the Dirac Hamiltonian includes a periodic Coulomb potential.
The wave functions are now periodic in time and quasiperiodic in space.
In this gauge, the wave function on the torus $0<x<2\pi$ has the form of a Bloch wave on a spatial lattice of unit cells $2\pi n<x<2\pi (n+1)$, with
$k$ being the Bloch wave momentum,
\begin{equation}
\Psi(x,k) = e^{ikx}u(x,k)
\end{equation}
In this way, a period in Euclidean time translates into a loop around the compact Brillouin zone. Just as in the case
of quantized charge transport in topological insulators, a nonzero Berry phase acquired by an eigenstate under adiabatic transport around the BZ describes
the transfer of charge across a unit spatial cell. Following standard arguments \cite{Resta}, we may describe this by a Berry connection constructed from the periodic part of the 
Bloch wave state,
\begin{equation}
\label{eq:connection}
A(k) = \Im\int_0^{2\pi}dx\; u^*(x,k)\frac{\partial}{\partial k}u(x,k)
\end{equation}
The topological order parameter is the closed Wilson loop around the BZ,
\begin{equation}
\label{eq:berryphase}
\theta = \oint dk A(k)
\end{equation}
Note that, as discussed in Ref. \cite{Thacker14}, the Berry connection is obtained from the phase of the periodic part $u(x,k)$ of the wave
function integrated over a single spatial cell of the lattice used to define the Bloch wave states. The Berry phase can thus be defined
locally as an order parameter by studying the spectral flow on a periodic unit cell of the lattice.
As we will show, the Berry phase is essentially determined by the flow of chiral charge in and out of the unit cell.

Topological transitions between discrete vacua with $\Delta\theta=\pm2\pi$ are represented by threading a unit of Berry flux through the loop in (\ref{eq:berryphase}).
To make this idea precise, we must analytically continue the Berry connection in complex momentum space and regard (\ref{eq:berryphase}) as a contour integral. 
For QED2, there is a natural analytic structure for $A(k)$ in complex momentum space, or more conveniently, in complex rapidity space,
where $k=m\sinh\xi$. The relevant analyticity properties of the wave functions are clarified by adding both a small mass term and a small
Thirring 4-fermion coupling to the Dirac action. Ultimately, we recover the spectral flow of the free massless Dirac Hamiltonian by taking the zero mass,
zero coupling constant limit of the massive Thirring model. The mass term provides a gap between the positive and negative energy bands, while the 4-fermion coupling 
provides a gap between the mass of the elementary boson and the fermion-antifermion threshold. Thus the eigenstates we use to analytically continue the Berry connection are
obtained from the Bethe ansatz solution of the massive Thirring model,
which we briefly review in the next section. The spectrum of this model can be described in terms of a single-quasiparticle energy function $\varepsilon(\xi)$, which
is a solution to the integral equation that follows from imposing periodic boundary conditions on the Bethe ansatz wave functions. 
The kernel of this integral equation is the derivative of the 2-body phase shift with respect to rapidity, 
\begin{equation}
K(\xi-\xi') = \frac{\partial \Delta(\xi-\xi')}{\partial\xi}
\end{equation}
which is a simple meromorphic function of complex rapidity. 
Physically, the spectral equations that follow from periodic boundary conditions describe the ``backflow'' of the negative energy Dirac sea 
in response to adiabatically changing the rapidity of a quasiparticle excitation. 
As we will show, the backflow integral defines a Berry phase for a given eigenstate in terms of a contour integral in the complex rapidity plane.
A discontinuity in the backflow integral arising from poles of $K$ represents the transfer of a charged fermion or antifermion to the boundary.
The analytic structure of the Berry connection constructed here also clarifies the role of bosonization and topological ordering from the Hamiltonian
spectrum viewpoint.

\section{Analytic structure of the massive Thirring spectrum near the free fermion point}

We consider the massive Thirring model (MTM) in 2 spacetime dimensions, defined as a massive Dirac fermion with a 4-fermion self interaction, described
by the Lagrangian
\begin{equation}
{\cal L} = \bar{\psi}\left(i\gamma^{\mu}\partial_{\mu}-m_0\right)\psi - \frac{1}{2}g_0j_{\mu}j^{\mu}
\end{equation}
where $j^{\mu}=\bar{\psi}\gamma^{\mu}\psi$. Under bosonization, this model is equivalent to the sine-Gordon (SG) model \cite{Coleman75}, the theory of
a real scalar field $\phi$ with Lagrangian
\begin{equation}
{\cal L} = \frac{1}{2\beta^2}\partial^{\mu}\phi \partial_{\mu}\phi - \frac{\alpha_0}{\beta^2}\left(\cos\phi - 1\right)
\end{equation}
Our discussion of the Berry connection will focus on the eigenstates of the MTM Hamiltonian
\begin{equation}
H = \int dx \left[-i\left(\psi_1^{\dag}\partial_x\psi_1 - \psi_2^{\dag}\partial_x\psi_2\right) + m_0(\psi_1^{\dag}\psi_2 +\psi_2^{\dag}\psi_1)
+2g_0\psi_1^{\dag}\psi_2^{\dag}\psi_2\psi_1)\right].
\end{equation}
The Bethe ansatz vacuum state is constructed on an empty Dirac sea by filling all of the negative energy states. If a particle state is excited above
the vacuum, its energy receives a contribution from the ``backflow'' of the Dirac sea. This results in an integral equation for the quasi-particle energy
function $\varepsilon(\xi)$ where $k=m\sinh\xi$ is the pseudomomentum variable of the excitation. The kernel of the integral equation that determines 
the spectrum is obtained from the 2-body phase shift which appears in the Bethe ansatz,
\begin{equation}
\Delta(\xi)=-i\log\left(\frac{\sinh(2i\mu-\xi)}{\sinh(2i\mu+\xi)}\right)
\end{equation}
Here, in standard notation, the Thirring coupling is parametrized by $\mu$,
\begin{equation}
\cot\mu = -\frac{1}{2}g_0
\end{equation}
The free fermion point is at $\mu=\pi/2$, and the attractive coupling region (where there are one or more fermion-antifermion 
bound states) is $\pi/2<\mu<\pi$.
The kernel which describes the backflow effect is given by the derivative of the 2-body phase shift,
\begin{equation}
\label{eq:kernel}
K(\xi) = \frac{d\Delta}{d\xi} = \frac{\sin2\mu}{\cosh\xi-\cos2\mu}
\end{equation}
This is a meromorphic function which is periodic in the imaginary rapidity direction and has simple poles with unit residue at
\begin{equation}
\xi=\pm 2i\mu + 2in\pi, \;\; n={\rm integer}
\end{equation}
Note that the poles in this function correspond to logarithmic branch points of the phase $\Delta(\xi)$.

To summarize a somewhat lengthy derivation \cite{Thacker81}, the particle spectrum of the model can be expressed
in terms of an analytic energy function for a quasiparticle excitation at complex rapidity,
\begin{equation}
\varepsilon(\xi) = \frac{1}{2}(\varepsilon_+(\xi)+\varepsilon_-(\xi))
\end{equation}
with
\begin{equation}
\label{eq:energy}
\varepsilon_{\pm}(\xi) = \int_{-\infty}^{\infty} K(\xi-(\alpha'+i\pi)) e^{\pm\gamma\alpha'} \frac{d\alpha'}{2\pi}
\end{equation}
where $\gamma=\frac{\pi}{2\mu}$, and the integral is along the real $\alpha'$ axis. (Here and elsewhere, we will use
$\alpha$ to denote a real rapidity and $\xi$ to denote a complex one.) The argument of the backflow kernel $K$ in (\ref{eq:energy})
is the relative rapidity between the mode $\xi$ and a vacuum mode on the $i\pi$ line at $\alpha'+i\pi$. 

For general attractive coupling the particle spectrum is
given by the construction of ``$n$-string'' excitations in the complex rapidity plane, which consist of n excitations 
arranged vertically in a bound-state configuration,
\begin{equation}
\label{eq:nstring}
\xi_i = \alpha +il(\pi-\mu),\;\; l=(n-1),(n-3),\ldots -(n-1)
\end{equation}
The energy of an $n$-string is just the sum of the $\varepsilon$'s for each of the $n$ modes. This reproduces the exact 
semiclassically quantized breather modes of the sine Gordon theory \cite{Thacker81}.
The values of $n$ are limited by the requirement that the $n$-string configuration remains within the region 
$-\pi <\Im{\xi} < \pi$, so near the free fermion point $\mu\approx \frac{\pi}{2}$, we have $n\leq 3$. The 1-string excitation turns out
to be the elementary sine Gordon boson. The 3-string disappears from the spectrum for $g_0\rightarrow 0$ and will not affect our considerations. 
On the other hand, the 2-string excitation remains in the spectrum even in the free fermion limit. The 2-string is central to our discussion of the 
Berry connection. It is a zero energy excitation 
which will prove to be the fermionic description of mod $2\pi$ jumps in the sine Gordon field. 

\begin{figure}
\vspace*{4.0cm}
\includegraphics{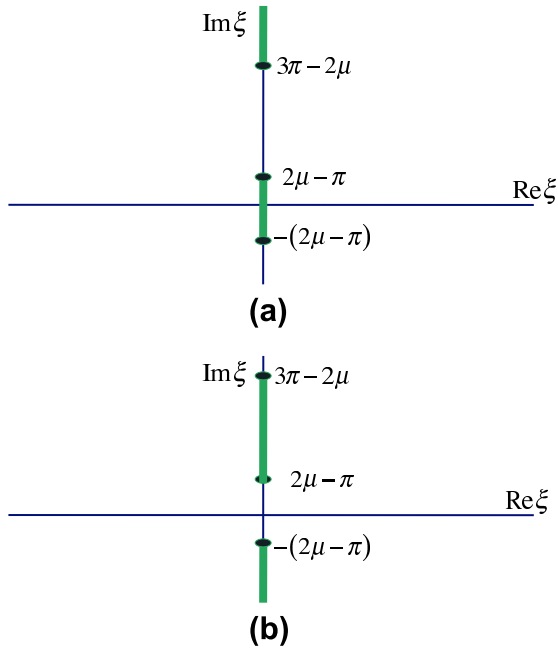}
\vspace{6.5cm}
\caption{(a) The fermionic branch choice for the 2-body phase shift. In the free fermion limit $\mu\rightarrow \pi/2$, the branch points at $\pm(2\mu-\pi)$ coalesce, the branch cuts vanish,
and the phase shift goes to zero on both sides of the imaginary $\xi$ axis. (b) The bosonic branch choice. Here the branch cuts vanish in the free boson limit $\mu\rightarrow \pi$. In the free
fermion limit, the phase shift reduces to a $2\pi$ step function across the imaginary $\xi$ axis.}
\label{fig:branchcuts}
\end{figure}

Although naively, an excitation at rapidity $\xi$ corresponds to a fermionic Bethe ansatz
mode, the phenomenon of ``hole-trapping'' \cite{Thacker81} has the effect that if $\xi$ is close enough to the real axis, the excitation is required to
have a hole on the $i\pi$ line attached to it, at the same value of $\Re{\xi}$. Thus, for $|\Im{\xi}|<2\mu-\pi$ the excitation is bosonic, and the energy given by (\ref{eq:energy}) 
is the energy of an elementary boson, as we show in the next section. However, if we analytically continue $\varepsilon(\xi)$ into the upper half $\xi$ plane, we hit an ionization threshold at
$\Im{\xi}=2\mu-\pi$, at which the function defined by (\ref{eq:energy}) changes suddenly when the pole at $\Im{\xi}=2\mu-\pi$ crosses the integration contour along $\alpha'$. 

It is convenient to define the fermion-antifermion phase shift,
\begin{equation}
\tilde{\Delta}(\xi)\equiv \Delta(\xi-i\pi)
\end{equation}
In the free fermion limit, the poles in 
\begin{equation}
\label{eq:boskernel}
\tilde{K}(\xi-\alpha') \equiv K(\xi-(\alpha'+i\pi)) = \frac{-\sin2\mu}{\cosh(\xi-\alpha')+\cos2\mu}
\end{equation}
at $\xi-\alpha'=\pm i(2\mu-\pi)$ pinch the real axis. There are two distinct branch choices for the
phase shift $\tilde{\Delta}(\xi)$, which we will refer to as ``fermionic'' and ``bosonic'' branch structures, shown in Fig. 1. The fermionic branch choice connects
the coalescing poles with a cut across the real axis. Note that for any finite Thirring coupling $\mu>\pi/2$, the two body phase shift is discontinuous along the real axis, but vanishes at large 
rapidity in both directions $\alpha'\rightarrow\pm\infty$.
Note also that, for the fermionic branch choice, the branch cuts disappear in the free fermion limit and the phase shift becomes identically zero. In order to construct the Berry
connection in complex rapidity, we must instead choose 
the bosonic branch structure shown in Fig. 1(b) where the branch cut does not cross the real axis, giving a phase shift that is continuous for real rapidity,
with no discontinuity at $\alpha=0$. In this case, the phase shift along the real axis does not reduce to zero in the free fermion limit, but to
zero or $\pm2\pi$, depending on the sign of the real part of the relative rapidity of the two modes.. 

\section{Bosonization and the Berry connection}

The bosonic choice of branch cuts for the phase shift $\tilde{\Delta}(\xi)$ allows us to regard this function as analytic in a strip around the real $\xi$ axis. It also 
provides a description of the Thirring/sine Gordon spectrum that is manifestly symmetric under charge conjugation. In the original Bethe ansatz solution of the model,
the states were built on an empty Dirac sea by first filling the negative energy modes, leading to a Dirac hole picture in which charge conjugation symmetry is
not manifest. Here instead, we start with the integral (\ref{eq:energy}), which defines the energy of bosonic excitations along the real axis and 
analytically continue to complex rapidity. Within the strip $|\Im{\xi}|<2\mu-\pi$, the excitation is bosonic. 
The energy of an excitation on or near the real rapidity axis can be calculated from the residues of the poles of $\tilde{K}$ in the periodic strip $0<\Im{\alpha'}<2\pi$. 
Integrating around the perimeter of the strip and exploiting the quasiperiodicity of the integrand in (\ref{eq:energy}), we get
\begin{equation}
\label{eq:contour}
\left(1-e^{2i\gamma\pi}\right)\varepsilon_{\pm}(\xi) = -\frac{1}{2}m_Fe^{\gamma(\xi-i\pi)}\left(1-e^{4i\gamma\pi}\right)
\end{equation}
giving
\begin{equation}
\varepsilon(\xi) = m_B\cosh\gamma\xi, \;\; |\Im{\xi}|<2\mu-\pi
\end{equation}
where $m_B=2m_F\cos\pi(1-\gamma)$ is the mass of the sine Gordon
boson. The two terms on the right hand side of (\ref{eq:contour}) are the residues of the poles in (\ref{eq:energy}) at $\alpha'=\xi+i(2\mu-\pi)$ and $\xi+i(3\pi-2\mu)$, respectively.
But when the energy function (\ref{eq:energy}) is continued into the upper half plane above the threshold line $\Im{\xi} = 2\mu-\pi$, the pole at $\xi+i(3\pi-2\mu)$ moves out of the
integration strip $0<\Im\alpha'<2\pi$, and the pole at $\xi+i(\pi-2\mu)$ enters it. Now if we integrate around the periodic strip, we get
\begin{equation}
\left(1-e^{2i\gamma\pi}\right)\varepsilon_{\pm}(\xi) = -\frac{1}{2}m_Fe^{\gamma(\xi-i\pi)}\left(1-e^{2i\gamma\pi}\right)
\end{equation}
and the integral gives the energy of a fermionic excitation,
\begin{equation}
\varepsilon(\xi) = -m_F\cosh\gamma(\xi-i\pi)
\end{equation}
In the free fermion limit $m_B\rightarrow 2m_F$ and $\gamma\rightarrow 1$, so for this case the energy is $\varepsilon(\xi)\rightarrow m_F\cosh\xi$. Thus, the mass of the particle represented by the energy function (\ref{eq:energy})
changes suddenly from $m_B\approx 2m_F$ to $m_F$ as the integration contour crosses the pole. This jump describes the ionization of the boson into its 
consituent fermion and antifermion. The mode at complex rapidity $\xi$ above this threshold represents a fermion. The antifermion that was bound to it below
the ionization threshold is represented by the residue of the pole at $\xi+i(3\pi-2\mu)$. This residue is no longer included in the energy of the bulk state defined by
(\ref{eq:energy}), so we take this to mean that the antifermion has gone to infinity or to a spatial boundary. In sine Gordon language, we have constructed a
kink by starting with the boson, which is a kink-antikink pair, and letting the antikink go to infinity to the right, as shown in Fig. 2(a). This physical argument exhibits the role of the
analytic structure of the fermion-antifermion kernel (\ref{eq:boskernel}) in defining a Berry connection. 

\begin{figure}
\vspace*{4.0cm}
\includegraphics{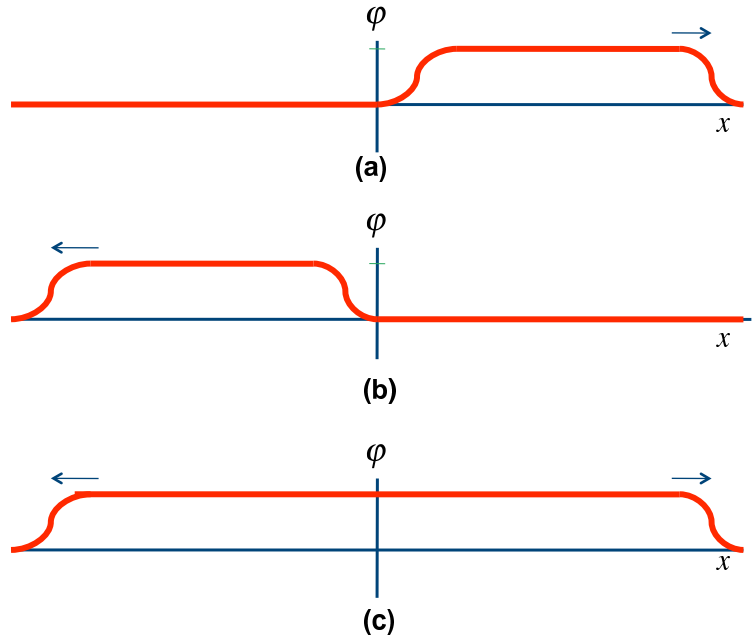}
\vspace{6.5cm}
\caption{(a) Constructing a fermion by ionizing a boson. (b) Constructing an antifermion by ionizing a boson. (c) A 2-string bound state of
the kink and antikink in (a) and (b).}
\label{fig:two_string1}
\end{figure}

\begin{figure}
\vspace*{4.0cm}
\includegraphics{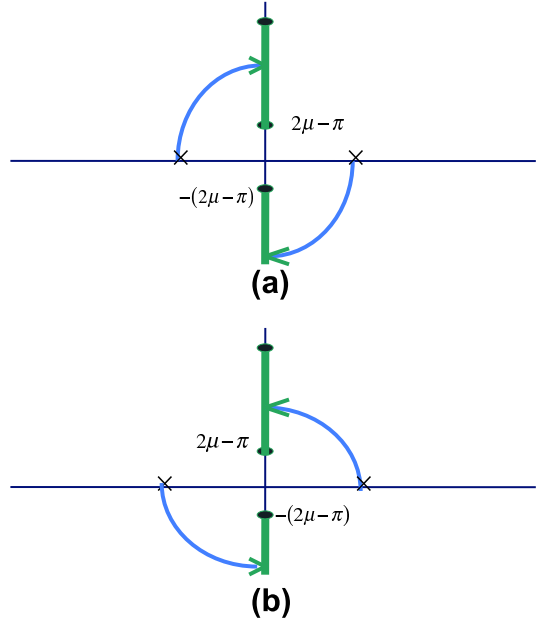}
\vspace{6.5cm}
\caption{(a) A left-handed 2-string with quantum nubers of a $\psi_1^{\dag}\psi_2$ excitation (b) A right-handed 2-string with quantum numbers of $\psi_2^{\dag}\psi_1$.}
\label{fig:two_string2}
\end{figure}

The cut structure that allows us to analytically continue the bosonic states in a strip around the real rapidity axis also restores manifest charge conjugation invariance
in the representation of the spectrum in the complex rapidity plane..
Analytic continuation into the lower half-plane exhibits the same ionization phenomenon for $-2\pi<\Im{\xi} < -(2\mu-\pi)$. But we may consistently interpret the fermionic
mode in this region of the lower half plane as an antifermion, with the residue of the pole at $\xi-(3\pi-2\mu)$ being the fermion, which has propagated to the boundary,
as shown in Fig. 2(b).
From the viewpoint of the fermionic Bethe ansatz,
we are constructing states in the lower half plane by a charge-conjugate ansatz in which the modes are antifermions and the holes are fermions.
In the bosonic strip along the real axis $-(2\mu-\pi)<\Im\xi<2\mu-\pi$, both the fermionic and the antifermionic Bethe ansatz are valid representations of the boson,
which map into each other by $CP$ conjugation. 
This transformation does not change the form of the 2-body phase shifts, so the identification of an $n$-string state as a vertical row of modes of the form (\ref{eq:nstring})
remains unchanged. In particular, there will still be zero energy states described by 2-strings. But the bosonic construction of the 2-string state clarifies
it's nature as a fermion-antifermion bound state and how it is related to mod $2\pi$ changes of the sine Gordon field. To see the essential point, consider constructing
a 2-string by starting with two bosonic modes on the real axis and analytically continuing one into the UHP and the other into the LHP. The mode in the UHP represents
a kink with an ionized antikink propagated off to the right, as shown in Fig. 2(a). The mode in the LHP represents an antikink, with an ionized kink propagated
off to the left, as shown in Fig. 2(b). The existence of the 2-string state tells us the quite plausible result that the kink and antikink in Figs. 2(a) and (b) can bind together to produce
a state with no spatial variation at all: the $\phi=2\pi$ vacuum, shown in Fig. 2(c). Note that, starting from the center of mass frame of the two bosons on the real axis,
there are two different ways to analytically continue them to form a 2-string, depicted in Fig. 3(a) and (b). The 2-string states obtained are related by CP conjugation
and carry opposite chiral charge. In terms of chiral fermion fields, the two types of 2-strings can be identified as $\psi_1^{\dag}\psi_2$ and $\psi_2^{\dag}\psi_1$ excitations.

In the absence of electromagnetic interactions, the vacua in which $\phi$ differs by mod $2\pi$ are degenerate. 
But if the EM coupling is turned on, the 2-strings become a bookkeeping device that keeps track of the charge polarization at a given location, and hence of the 
number of units of electric flux at that point. If a left moving 
fermion or a right moving antifermion passes a point, $\phi$ increases by $+2\pi$, while a right moving fermion or left moving antifermion decreases $\phi$ by $2\pi$.
This can be understood in terms of our Berry phase construction by considering
the spectral flow argument given in Section I and in Ref. \cite{Thacker14}. The Bethe ansatz integral equation that defines the spectrum in terms of the backflow integral
(\ref{eq:energy}) arises from imposing periodic boundary conditions on a box of length $L$. If periodicity is imposed on the many-body coordinate space wave function, the momentum
variables $k_i = m_0\sinh\xi_i$ in the Bethe ansatz must satisfy 
\begin{equation}
\label{eq:PBC}
e^{im_0L \sinh\xi_i}e^{i\sum_{j\neq i}\tilde{\Delta}(\xi_i-\xi_j)} = 1
\end{equation}
When applied to a mode $k_i$ in the negative energy sea (on the $i\pi$ line), the boundary condition (\ref{eq:PBC}) becomes the integral equation for the vacuum backflow 
in the $L\rightarrow\infty$ limit.  
We see that the periodic boundary conditions include, in addition to the quasiparticle momentum phase $e^{ik_iL}$, also a phase factor from 2-body phase
shifts with all the other particles in the state. For the vacuum state, the $k_j$'s consist of occupied negative energy modes along the $i\pi$ line. The backflow 
associated with an excitation above the vacuum is determined by the relative phase between the excited state and ground state in (\ref{eq:PBC}). 

This shows directly how the Bethe ansatz spectral equations serve to define a Berry connection for Hamiltonian eigenstates. 
When a background EM field is introduced, 
the Hamiltonian spectral flow as a function of the parameter
$k$ in (\ref{eq:spectralflow}) is defined by the spectral equations (\ref{eq:PBC}) for $k=k_i$. After taking the ratio of the excited state to the ground state PBC's, the only
phase shifts which remain in Eq. (\ref{eq:PBC}) (in the free fermion limit) are those associated with excitations. 
If we had chosen the fermionic
branch structure, the 2-body phase shifts would all vanish in the free fermion limit, and (\ref{eq:PBC}) would give the spectrum for noninteracting fermions. But with a small (ultimately 
vanishing) Thirring coupling and the bosonic choice of branch cuts, the phase shift in the free fermion limit 
becomes a step function which jumps by $\pm2\pi$ when the real part of the ``probe'' rapidity $\xi=\xi_i$ crosses the rapidity of the excitation $\xi_j$. The Berry
phase (\ref{eq:berryphase}) for the probe particle going all the way around the BZ is just the sum of $\pm2\pi$ phase shifts over the other excitations in the box.
As discussed in the Introduction and in \cite{Thacker14}, the adiabatic time development given by the Dirac equation with a background gauge field $A_1=Ft\equiv k$ provides a
definition of the transport in $k$-space which determines the Berry connection.

Our construction of a Berry connection for QED2 is quite similar to the theory of polarization in topological insulators \cite{Vanderbilt}. In fact, it is easy to see that bosonization
in QED2 is just a description of the fermionic currents in terms of a local polarization, the sine-Gordon field. When periodic boundary conditions are imposed on the Bethe
ansatz wave functions, we single out a particular quasiparticle momentum ($k_i$ in (\ref{eq:PBC})) and treat it as a Bloch wave momentum. The 2-body phase shift factors in 
(\ref{eq:PBC}) play the role of $u(x,k)$, the periodic part of the Bloch wave. As in topological insulator theory, the Berry phase is associated with topologically nontrivial transport
of $u(x,k)$ in momentum space. The full Berry connection is given by the analytic structure in complex rapidity space, with the poles of the analytically continued
2-body phase shift representing topological
transitions in which a neutral boson ionizes into an unbound fermion-antifermion pair.

\section{Discussion}

In the bosonization of the Dirac current in QED2, $j_{\mu}=\varepsilon_{\mu\nu}\partial^{\nu}\phi/2\pi$ 
the boson field can be interpreted as the electric polarization, which, in a dielectric medium, is
related to the current by $\vec{\nabla}\cdot \vec{P} = j_0,\;\partial_t \vec{P}= \vec{j}$. Thus in 2D, the local
rate of change of the polarization is just the axial charge density,,
\begin{equation}
j^5_0 = \frac{1}{2\pi}\partial_0 \phi
\end{equation}
In the previous Section, we saw that, in the Bethe ansatz framework, the Berry phase arises from the additional phase acquired by momentum transport of a quasiparticle 
due to the 2-body phase shifts from the other occupied modes in the state. This is related to the spectral flow of the Dirac Hamiltonian by the fact that the
Berry phase appears in the periodic boundary conditions (\ref{eq:PBC}) which determine the spectrum. In the free fermion limit, these 2-body phase shifts reduce to
step functions which jump by $\pm 2\pi$ when the real part of the relative rapidity $\Re{(\xi_i-\xi_j)}\equiv\alpha_i-\alpha_j$ changes sign. The sign of the jump in the
2-body phase shift is determined by the chirality of the excitation crossed, i.e. in the massless limit, whether it is a $\psi_1^{\dag}$ or a $\psi_2^{\dag}$ excitation.
The only excitations we need to consider are 1-strings and 2-strings. The 1-string state is a bound state of a fermion and an antifermion which have the exactly the
same value of $\Re\xi$. So the two phase shifts have opposite sign and cancel in the Berry phase. 
On the other hand, for a 2-string, the phase shifts add,
and there is a mod $2\pi$ contribution to $\theta$, representing the ionization and flow of a unit of chiral charge to the boundaries.
Thus, in the free fermion limit (with the bosonic branch choice) the sum of 2-body phase shifts in (\ref{eq:PBC}) just 
calculates the net chiral charge in the box. We conclude that Berry phase on a unit cell can be identified with the sine Gordon field averaged over the cell,
\begin{equation}
\label{eq:deltheta}
\Delta \theta = \frac{1}{2\pi}\Delta\int_0^{2\pi} \phi dx = \int_0^Tdt\int_0^{2\pi}\frac{dx}{2\pi} j_0^5 = \int_0^TQ_0^5dt
\end{equation}
The last expression shows that the Berry phase is a chiral phase rotation generated by the total axial charge on the unit cell.
The interpretation of the Berry phase as a chiral vacuum $\theta$ angle generalizes to 4-dimensional QCD \cite{Thacker14}.

The utility of the bosonic choice of branch structure in defining the Berry phase is that an elementary boson (i.e. a 1-string + hole excitation) does not contribute to the Berry phase.
In a 1-string, the mode on the real axis and its attached hole are both at the same value of $\Re{\xi}$, so in the free fermion limit, the two phases cancel in the overall
sum of phases in (\ref{eq:PBC}). On the other hand,
a 2-string is formed by two bosonic excitations analytically continued into the upper and lower rapidity planes. The contributions to the Berry phase add instead of cancelling,
reflecting the fact that, in the formation of a 2-string, there has been a net flow of chiral charge to the boundaries. 

This work was supported by the Department of Energy under grant DE-SC00079984.

\end {document}